\documentclass[preprintnumbers,article,amsmath,amssymb,floatfix,10pt,prd,onecolumn,
superscriptaddress,nofootinbib]{revtex4}
\usepackage[colorlinks=true, pdfstartview=FitV, linkcolor=blue, citecolor=red, urlcolor=magenta]{hyperref}
\usepackage{bbm}
\usepackage{amsfonts}
\usepackage{mathrsfs}
\usepackage{latexsym}
\usepackage{epsfig}
\usepackage{epstopdf}
\usepackage{epstopdf}
\usepackage{graphicx}
\usepackage{amssymb}
\usepackage{amsmath}
\usepackage{dcolumn}
\usepackage{bm}
\usepackage{float}
\usepackage{color}
\usepackage{comment}
\usepackage{xcolor}
\begin{document}
\title{\bf Constraining study of charged gravastars solutions in symmetric teleparallel gravity}

\author{Faisal Javed}
\email{faisaljaved.math@gmail.com}
\affiliation{Department of
Physics, Zhejiang Normal University, Jinhua 321004, People's
Republic of China}
\affiliation{Zhejiang Institute of
Photoelectronics and Zhejiang Institute for Advanced Light Source,
Zhejiang Normal University, Jinhua, Zhejiang 321004, China}
\affiliation{Institute of Fundamental and Applied Research, National Research University TIIAME, Kori Niyoziy 39, Tashkent 100000, Uzbekistan}
\author{Arfa Waseem}
\email{arfa.waseem@gcwus.edu.pk} 
\affiliation{Department of
Mathematics, Government College Women University, Sialkot, Pakistan}

\author{G. Mustafa}
\email{gmustafa3828@gmail.com (Corresponding Author)} 
\affiliation{Department of
Physics, Zhejiang Normal University, Jinhua 321004, People's
Republic of China}
\affiliation{Institute of Fundamental and Applied Research, National Research University TIIAME, Kori Niyoziy 39, Tashkent 100000, Uzbekistan}

\author{Fairouz Tchier}
\email{ftchier@ksu.edu.sa} \affiliation{Mathematics Department, College of Science, King Saud University, P.O. Box 22452, Riyadh 11495, Saudi Arabia}

\author{Farruh Atamurotov}
\email{atamurotov@yahoo.com}
\affiliation{New Uzbekistan University, Movarounnahr street 1, Tashkent 100000, Uzbekistan}
\affiliation{Institute of Fundamental and Applied Research, National Research University TIIAME, Kori Niyoziy 39, Tashkent 100000, Uzbekistan}
\affiliation{Tashkent State Technical University, Tashkent 100095, Uzbekistan}

\author{Bobomurat Ahmedov}
\email{ahmedov@astrin.uz}

\affiliation{Institute of Fundamental and Applied Research, National Research University TIIAME, Kori Niyoziy 39, Tashkent 100000, Uzbekistan} 
\affiliation{Ulugh Beg Astronomical Institute, Astronomy St 33, Tashkent 100052, Uzbekistan} 
\affiliation{Institute of Theoretical Physics, National University of Uzbekistan, Tashkent 100174, Uzbekistan}

\author{Ahmadjon Abdujabbarov}
\email{ahmadjon@astrin.uz}

\affiliation{Ulugh Beg Astronomical Institute, Astronomy St 33, Tashkent 100052, Uzbekistan} 
\affiliation{Institute of Theoretical Physics, National University of Uzbekistan, Tashkent 100174, Uzbekistan}
\affiliation{University of Tashkent for Applied Sciences, Str. Gavhar 1, Tashkent 100149, Uzbekistan}

\begin{abstract}
This study explores the effect of charge on a special astronomical object known as a gravastar, which is viewed as an alternative to a black hole. Based on the conjecture put out by Mazur and Mottola in general relativity, the study primarily focuses on the consequences of $f(Q)$ gravity. The internal domain, the intermediate shell, and the external domain are the three separate sections that make up a gravastar. Using a particular $f(Q)$ gravity model that includes conformal Killing vectors to analyze these areas, we discover that the inner domain shows a repulsive force on the spherical shell since it is assumed that pressure is equivalent to negative energy density. The intermediate shell is made up of ultrarelativistic plasma and pressure, which is proportional to energy density and balances the repulsive force from the interior domain. For the exterior region, we use two approaches first we calculate the vacuum exact solution, and second, consider the Reissner-Nordström metric. Then, we match these spacetimes through junction conditions and explore the stability constraints for both cases. Our results show that charged gravastar solutions with non-singular physical parameters including length, energy, entropy, and equation of state parameter are physically realistic.

\end{abstract}

\maketitle

\date{\today}

\section{Introduction}\label{sec:1}

In recent years, gravitational collapse as well as stellar
configuration are major astrophysical phenomena that have captured
the interest of many researchers. White dwarfs, neutron stars, and
black holes (BHs) are examples of new and massive compact celestial
bodies that are produced as the consequence of gravitational
collapse. Currently, several authors have suggested that the
most dense celestial objects besides the BHs could be created by
the gravitational collapse of a massive star. In this respect, by taking
into account an expanded concept of Bose-Einstein condensation in
the gravitational system, Mazur and Mottola \cite{1} first proposed
a novel theory of collapsing stellar objects dubbed as gravitational
vacuum star (gravastar). It is considered that the gravastar model
offers a solution to the problems with classic BHs and fulfills all
theoretical requirements for a stable stellar configuration. In this
conjecture, it is anticipated that quantum vacuum fluctuations
contribute significantly to the dynamics of the collapse. A phase
transition occurs leading to a repulsive de-Sitter core that
balances the collapsing object and prevents the production of
the horizon as well as the singularity \cite{2}. Such transformation
takes place extremely near to the bound $\frac{2m}{r}=1$,
making it very challenging for an outside observer to identify the
gravastar from a genuine BH.

The gravastar geometry is illustrated by three distinct zones where
the internal geometry ($r\geq 0, r< \mathcal{R}_{1}$) is based upon
the isotropic de Sitter core comprising the equation of state (EoS)
$-\rho=p$. The external vacuum zone ($r> \mathcal{R}_{2}$) is
demonstrated by the Schwarzschild geometry with EoS $p=\rho=0$. The
internal and external regions are segregated through a thin shell
($\mathcal{R}_{1}<r< \mathcal{R}_{2}$) of stiff fluid having EoS
$\rho=p$ in which $\mathcal{R}_{1}$ and $\mathcal{R}_{2}$ depict the
internal and external radii of gravastar. After the proposal of
Mazur and Mottola, an enormous discussion has been done on gravastar.
Visser and Wiltshire \cite{3} analyzed the stable structure of
gravastars and examined that distinct EoSs yield the stable
configuration of gravastars. Carter \cite{4} presented novel exact
gravastar solutions and observed the impact of EoS on different
zones of gravastar geometry. Bili\'{c} et al. \cite{5} obtained the
gravastars solutions by considering the Born-Infeld phantom instead
of de Sitter spacetime and observed that at the star's core, their
findings can manifest the dark compact configurations.

Horvat and Iliji\'{c} \cite {6} investigated the role of energy
bounds in the intermediate domain of the gravastar and checked the
stable configuration via radial perturbations as well as the speed of
sound. Several authors \cite{7}-\cite{12} displayed the internal
structure of gravastar by adopting distinct EoSs concerning
different aspects. Lobo and Arellano \cite{13} constructed various
gravastar structures with the inclusion of nonlinear electrodynamics
and discussed some specific features of their structures. Horvat et
al. \cite{14} extended the concept of gravastar by inducing an electric
field and inspected the stability of the internal and external
regions. On the same ground, Turimov et al. \cite{15} observed the
consequences of magnetic field on the gravastar geometry and
determined the exact solutions of rotating gravastar. The detailed study of thin-shell developing from different inner and outer manifold as studied in \cite{fk1}-\cite{fk6}.  Recently, Javed and Lin \cite{ff1} studied the exact gravastar solutions in general relativity by considering the effects of a cloud of strings and quintessence field. 

In cosmological conjectures, the causes behind the cosmological
expansion yields the formation of modified gravitational theories to
general relativity (GR). In these theories, the notion of matter and
curvature coupling yields distinct approaches like $f(R,T)$ gravity
\cite{16}, $f(R,T, R_{\alpha\beta}T^{\alpha\beta})$ gravity
\cite{17} and $f(\mathcal{G},T)$ theory \cite{18} in which $R$
denotes the curvature invariant, $T$ depicts the trace of
energy-momentum tensor (EMT) and $\mathcal{G}$ describes the
Gauss-Bonnet invariant. There also exists some modified theories other than these theories and a lot of different scenarios have been discussed in these modified theories. Zubair et al. \cite{18a} discussed the thermodynamics of some newly developed $f(R,L_{m}, T)$ theories with conserved energy-momentum tensor. Zubair and Farooq \cite{18b} presented the reconstruction as well as some dynamical aspects of bouncing scenarios in $f(T,\mathcal{T})$ gravity.

The exploration of gravastar geometries has
prompted the astrophysicist to analyze the influences of extended
gravitational conjectures on different modes of gravastar. In
$f(R,T)$ background, Das et al. \cite{19} discussed the gravastar
geometry and observed its characteristics graphically corresponding
to various EoSs. Shamir and his co-author \cite{20} depicted the
non-singular gravastar solutions and derived mathematical
formulations of various physical terms in $f(\mathcal{G},T)$
framework. Sharif and Waseem \cite{21} inspected the effects of
Kuchowicz metric potential on gravastar geometry in $f(R,T)$ theory.
Yousaf and his collaborators \cite{22,23} examined the features of
gravastar concerning some specific constraints in different
theories. Bhatti et al. \cite{24} demonstrated some particular
features of gravastar geometry with a specific EoS in
$f(R,\mathcal{G})$ scenario. Sharif and Naz \cite{25,26} studied the
salient aspects of gravastar in the absence and presence of electric
field in $f(R,T^{2})$ gravity. Azmat et al. \cite{26a} developed the analytic form of gravastar with anisotropic and non-uniform characteristics through the gravitational decoupling procedure in $f(R,T)$ gravity. The inclusion of charge parameters plays an important role in discussing the solutions of ultracompact objects. Tello-Ortiz et al. \cite{26b} obtained the new hairy spherically symmetric and asymptotically flat-charged BH solution starting from the Reissner-Nordstr\"{o}m manifold by means of extended geometric deformation approach. Azmat and Zubair \cite{26c} analyzed the impact of the charge on the anisotropic non-uniform gravastar model in $f(R,T)$ framework.

An inheritance symmetry with a collection of conformal Killing
vectors (CKVs) is crucial for determining a natural consistent
connection between the components of geometry and matter for giant
celestial bodies via field equations. In contrast to previous
analytical methods, these vectors are used to get precise analytical
solutions in more suitable forms. Implementing these vectors, the
large system of nonlinear partial differential equations can be
converted into an ordinary one. Usmani et al. \cite{27} discussed
the different aspects of gravastar with electric charge and obtained
solutions for its distinct eras associated with CKVs. Sharif and
Waseem \cite{28} investigated the impact of the charge on gravastar in
$f(R,T)$ theory by considering these vectors. Bhar and Rej \cite{29}
presented the charged gravastar model conceding the CKV in the
framework of $f(\mathcal{T})$ theory. Sharif and his collaborators
\cite{30,31} explored the analytic solutions of uncharged as well as
charged gravastar models by employing CKVs in $f(R,T^{2})$ theory. The study of thin-shell gravastars developed from inner de Sitter and outer various BH geometries is presented in  \cite{fa7}-\cite{fa12}.

Recently, the $f(Q)$ gravity \cite{32} as an extended gravitational
theory gained a lot of attention. It falls under the category of
symmetric teleparallel gravity (STG) in which gravity is linked to
the non-metricity and $f(Q)$ displays the generic function of the
non-metricity scalar $Q$. The basic difference between GR and
teleparallel gravity is the part of an affine connection instead of
the physical manifold. The accelerated cosmic expansion is characterized
by $f(Q)$ gravity at least to the same statistical precision as the
other well-known extended gravitational theories \cite{33}.
Different aspects like energy conditions \cite{34} and Newtonian
limit \cite{35} have been discussed in this context. Hassan et al.
\cite{36} inspected the solution of wormholes (WHs) by considering
the Lorentzian as well as the Gaussian configurations of exponential
and linear models. Mustafa et al. \cite{37} determined the WH
solutions in this framework with the Karmarkar condition. They presented
that such conditions yield the possible existence of traversable WHs
obeying the energy bounds. Sharma and his collaborators \cite{38}
explored the WH solutions in the light of STG. They observed that
some models of the $f(Q)$ gravity by employing the exact shape and
redshift function may generate such solutions that satisfy the
energy constraints. The detailed study of wormhole structure via thin-shell in different modified theories of gravity is presented in \cite{fa1}-\cite{fa6}. Also, the study of observational constraints in the framework of accelerated emergent $f (Q)$ gravity model is explored in  \cite{fa13}. D'Ambrosio et al. \cite{39} obtained the static spherically symmetric BH solutions in this gravity. Javed and his
collaborators \cite{40,41} examined the thermodynamics of perturbed
BHs as well as charged BHs in this framework. Classical works comprising the modified theories of gravity on compact objects by different authors can be found
in the literature \cite{Jr1,Jr2,Jr3,Jr4,Jr5,Jr6,Jr7,Jr8,Jr9,Jr10,Jr11}.

In this manuscript, we construct the non-singular solutions of
charged gravastar in the light of $f(Q)$ gravity with CKV. This work explores the effect of charge on gravastars, a novel astronomical object proposed as a black hole substitute. Based on the hypothesis developed by Mazur and Mottola in the context of general relativity, this study focuses on the implications of $f(Q)$ gravity. Three separate domains make up the gravastar: the exterior, intermediate shell, and internal sections. Using a particular $f(Q)$ gravity model with conformal Killing vectors, the paper shows that pressure is equal to negative energy density and that the inner domain pushes against the spherical shell. The intermediate shell balances the repulsive force from the interior domain with ultrarelativistic plasma and pressure proportionate to energy density. Two methods are used in the exterior region: first, the vacuum precise solution is calculated, and then the Reissner-Nordström metric is taken into account. We investigate stability requirements for both scenarios and match these spacetimes through junction conditions. The results imply that charged gravastar solutions are physically realistic. The
manuscript is managed in the following pattern. The next section is
dedicated to presenting the fundamentals of $f(Q)$ gravity. Section
\textbf{3} manifests the physical aspects of charged gravastars with
CKVs. The physical characteristics like the EoS parameter and proper
length of the intermediate thin shell region are discussed in section
\textbf{4}. The outcomes of this manuscript are elaborated in the
last section.

\section{Fundamentals of $f(Q)$ gravity}\label{sec:2}

In STG, we assume that the gravastar being studied is
presented within a differentiable Lorentzian manifold known as
$\mathcal{M}$. This manifold can be sufficiently described by the
metric $g_{ij}$, the determinant $g$ and the affine connection
$\Gamma$ which is defined as:
\begin{equation}
g=g_{ij}dx^i\otimes dx^j,
\end{equation}
Herein $\Gamma^\epsilon_{\,\,\,\,\varepsilon}$ indicates the
connection that can be regenerated in the form of disformation,
contortion tensor and Levi-Civita \cite{42}:
\begin{equation}
\Gamma^{\epsilon}_{\,\,\,\,\varepsilon}=w^{\epsilon}_{\,\,\,\,\varepsilon}+K^{\epsilon}_{\,\,\,\,\varepsilon}+L^{\epsilon}_{\,\,\,\,\varepsilon}.
\end{equation}
The above-said relation can be rewritten as follows:
\begin{equation}
\Gamma^{\epsilon}_{\,\,\,\,ij}=\gamma^{\epsilon}_{\,\,\,\,ij}+K^{\epsilon}_{\,\,\,\,ij}+L^{\epsilon}_{\,\,\,\,ij}.
    \label{eq:2}
\end{equation}
In the above Eq. (\ref{eq:2}), the expressions $L$, $\gamma$, and
$K$, define the disformation tensors, affine connection, and
contortion respectively. Now, the non-metricity term and it's associated
tensor in $f(Q)$ background are provided by
\begin{equation}
Q^\epsilon_{\,\,\,\,\varepsilon}=\Gamma_{(ab)},\quad Q_{\epsilon
ij}=\nabla_\epsilon g_{ij}.
\end{equation}
In the aforementioned relation, the tensor symmetric part satisfies
the relation is given by
\begin{equation}
F_{(ij)}=\frac{1}{2}\bigg(F_{ij}+F_{j i}\bigg).
\end{equation}
It is noted from the above discussion that if contortion disappears,
then only the disformation tensor is expressed as:
\begin{equation}
    Q_{\epsilon ij}=-L^{\varepsilon}_{\,\,\,\,\epsilon i}g_{\varepsilon j}-L^{\varepsilon}_{\,\,\,\,\epsilon j}g_{\varepsilon i},
\end{equation}
The disformation tensor is defined by the following formula:
\begin{equation}
    L^{\epsilon}_{ij}=\frac{1}{2}Q^{\epsilon}_{ij}-Q_{(ij)}^{\epsilon}.
    \label{eq:3}
\end{equation}
In the STG formalism, the non-metricity scalar within the scope of
superpotential is provided as
\begin{equation}
    Q=-P^{\epsilon ij}Q_{\epsilon ij},
\end{equation}
where $-P^{\epsilon ij}$ define the superpotential, which is
revealed as:
\begin{equation}
    P^{\epsilon}_{\,\,\,\,ij}=\frac{1}{4}\bigg[2Q^{\epsilon}_{\,\,\,\,(ij)}-Q^{\epsilon}_{\,\,\,\,ij}+Q^\epsilon g_{ij}-\delta^{\epsilon}_{(i}Q_{j)}-\overline{Q}^\epsilon g_{ij}\bigg].
\end{equation}
Here, $Q^{\epsilon}=Q^j_{\,\,\,\,\epsilon j}$ and
$\overline{Q}_\epsilon=Q^i_{\,\,\,\,\epsilon i}$ are important
unconstrained traces for $Q_{\epsilon ij}=\nabla_\epsilon g_{ij}$.
After the necessary calculations, we have the following modified
action integral for $f(Q)$ gravity \cite{43} for STG formalism:
\begin{equation}
    \mathcal{S}=\int d^4x \sqrt{-g}f(Q)+[\mathcal{S}_{\mathrm{M}}+L_{e}].
    \label{eq:3}
\end{equation}
Here, $\mathcal{S}_{\mathrm{M}}$ defines the action integral and
$L_{e}$ is a candidate for electric field. The modified field
equations for $f(Q)$ gravity can be obtained by varying the Eq.
(\ref{eq:3}), which are calculated as:
\begin{equation}
\frac{2}{\sqrt{-g}}\nabla_\gamma\left(\sqrt{-g}\,f_Q\,P^\gamma\;_{ij}\right)+\frac{1}{2}g_{ij}f
+ f_Q\left(P_{i\gamma i}\,Q_j\;^{\gamma i}-2\,Q_{\gamma i
i}\,P^{\gamma i}\;_j\right)=-T_{ij}, \label{eq:12}
\end{equation}
where $f_Q\equiv\frac{df}{dQ}$ and $T_{ij} $ represents the energy
momentum tensor, which is given as:
\begin{equation}
    T_{ij}+E_{ij}=-\frac{2}{\sqrt{-g}}\frac{\delta(\sqrt{-g} \mathcal{L}_{\mathrm{M}})}{\delta g^{ij}}.
\end{equation}
Here $\mathcal{L}_{\mathrm{M}}$ manifests the Lagrangian density of
matter field and it obeys the identity $\int \sqrt{-g}
\mathcal{L}_{\mathrm{M}}d^4x=\mathcal{S}_{\mathrm{M}}[g,\Gamma,\Psi_i]$.
Further, on changing the action within the scope of affine
connection $\Gamma^\epsilon_{\,\,\,\,ij}$, we obtain the following
relation:
\begin{equation}
\nabla_i \nabla_j \left(\sqrt{-g}\,f_Q\,P^\gamma\;_{ij}\right)=0.
\label{eq:14}
\end{equation}
Corresponding to the spherically symmetric metric, the line element
is provided as:
\begin{equation}\label{eq:13a}
 ds^2= e^{a} dt^2-e^{b} dr^2-r^2(d\theta^2+ sin^2 \theta d\phi^2).
\end{equation}
Further, for isotropic matter configuration, the energy-momentum
tensor becomes:
\begin{equation}\label{eq:14a}
T_{ij}=diag(e^{a}\rho, e^{b}p,r^{2} p, r^2 p sin^{2}\theta).
\end{equation}
Now the other involved tensors like electromagnetic stress energy
tensor $E_{i \upsilon}$ is given by
\begin{equation}\label{eq:14b}
E_{i j}=2(F_{i \zeta}F_{j \zeta}-\frac{1}{4}g_{i j}F_{\zeta
\chi}F^{\zeta \chi}),
\end{equation}
where
\begin{equation*}
F_{i j}=\mathcal{A}_{i, j}-\mathcal{A}_{j, i}.
\end{equation*}
Further, $F_{i j}$ denotes the tensor for electromagnetic field
provided by
\begin{eqnarray}
F_{i j,\zeta}+F_{\zeta i,j}+F_{j \zeta,i}=0,\label{22}\\
(\sqrt{-g}F^{i j})_{,j}=\frac{1}{2}\sqrt{-g} j^{i}.\label{23}
\end{eqnarray}
The electromagnetic four potentials are denoted by
$\mathcal{A}_{i}$, and all components of electromagnetic field
tensor become zero except the radial component $F_{01}$. Equation
(\ref{23}) is fulfilled when $F_{01}$ is antisymmetric, i.e.,
$F_{01}=-F_{10}$. Equation (\ref{eq:14b}) can be used to obtain the
electric field ($E$).
\begin{equation}\label{24}
E(r)=\frac{1}{2r^2}e^{a(r)+b(r)}\int_{0}^{r}\sigma(r)e^{b(r)}r^{2}dr=\frac{q(r)}{r^{2}},
\end{equation}
where $q(r)$ and $\delta$ manifest the total charge and charge
density of the system, respectively. Now, we will derive the field
Eqs. (\ref{eq:12}) and (\ref{eq:14}) for spherically symmetric
spacetime, which is calculated as:
\begin{eqnarray}
     8\pi T_{tt}&& = \frac{e^{a-b}}{2r^2}  [2rf_{QQ} Q'(e^{b} -1)+ f_{Q}[(e^{b} -1)(2+ra')+(1+e^{b})r b']+fr^{2}e^{b}],\label{eq:15}\\
      8\pi T_{rr}&& =- \frac{1}{2r^2}  [2rf_{QQ} Q'(e^{b} -1)+ f_{Q}[(e^{b} -1)(2+ra'+rb')-2ra']+fr^{2}e^{b}],\label{eq:16}\\
      8\pi T_{\theta\theta} &&= - \frac{r}{4e^{b}}  [-2rf_{QQ} Q' a' + f_{Q}[2a'(e^{b} -2) -ra'^{2}+b'(2e^{b}+ra')-2ra'']+2fre^{b}],\label{eq:17}
\end{eqnarray}
Now, by using the Eqs. (\ref{eq:14a}) and (\ref{eq:14b}) into
Eqs. (\ref{eq:15}-\ref{eq:17}), the modified form of $f(Q)$ field
equation becomes

\begin{eqnarray}
    8\pi\rho +E^{2} &&= \frac{1}{2r^2 e^{b}}  [2rf_{Q Q} Q'(e^{b} -1)+ f_{Q}[(e^{b} -1)(2+ra')+(1+e^{b})r b']+fr^{2}e^{b}],\label{eq:18}\\
    8\pi p -E^{2} &&= - \frac{1}{2r^2 e^{b}}  [2rf_{Q Q} Q'(e^{b} -1)+ f_{Q}[(e^{b} -1)(2+ra'+rb')-2ra']+fr^{2}e^{b}],\label{eq:19}\\
    8\pi p +E^{2} &&= - \frac{1}{4re^{b}}  [-2r f_{Q Q} Q' a' + f_{Q}[2a'(e^{b} -2)-  ra'^{2}+b'(2e^{b}+ra')-2ra'']+2f re^{b}],\\\label{eq:20}
    \sigma &&= \frac{e^{-b (r)} \frac{\partial \left(r^2 E \right)}{\partial r}}{4 \pi  r^2}.\label{eq:21}
\end{eqnarray}
in which the non-metricity scalar is illustrated by
\begin{equation}\label{eq:22}
    Q =\frac{1}{r}(a'+b')(e^{-b}-1).
\end{equation}
For the current analysis, the linear model of $f(Q)$ gravity is
taken into account which is exhibited as:
\begin{equation}\label{eq:23}
f=\alpha  Q+\Phi,
\end{equation}
where $\alpha$ and $\Phi$ are the model parameters. Now, by plugging
Eq. (\ref{eq:22}) and Eq. (\ref{eq:23}) into Eqs.
(\ref{eq:18}-\ref{eq:20}), one can get the following revised final
version of field equations:
\begin{eqnarray}
    8\pi\rho +E^{2} &&= \frac{2 \alpha +2 \alpha  e^{-b(r)} \left(r b'(r)-1\right)+r^2 \Phi }{2 r^2},\label{eq:24}\\
    8\pi p -E^{2} &&= -\frac{-2 \alpha  e^{-b(r)} \left(r a'(r)+1\right)+2 \alpha +r^2 \Phi }{2 r^2},\label{eq:25}\\
    8\pi p +E^{2} &&= \frac{e^{-b(r)} \left(2 \alpha  r a''(r)+\alpha  \left(r a'(r)+2\right) \left(a'(r)-b'(r)\right)-2 r \Phi  e^{b(r)}\right)}{4 r},\\\label{eq:26}
\end{eqnarray}

\section{Physical quantities of charged gravastar with Conformal Motion}

The study of charged gravastars in $f(Q)$ gravity is an important endeavor that combines fundamental motives in gravitational physics with the discovery of unique astronomical phenomena. The study's major goal is to understand the complex interplay between gravastars' charged nature and the changes introduced to the gravitational sector by $f(Q)$ gravity. This project is motivated by the theoretical richness of $f(Q)$ gravity, which goes beyond general relativity by including a function $f(Q)$ of the non-metricity scalar $Q$. Symmetries are very important to developing the inherent connection
between geometry and matter through Einstein's equations. The CKVs
among the well-known symmetries, offer superior outcomes as they
provide a more profound understanding of geometry. The equation
that governs CKVs is defined as:
\begin{eqnarray}\label{Killing}
\pounds_\xi g_{ij}=\xi~ g_{ij},
\end{eqnarray}
For a four-dimensional metric, where $i$ and $j$ range from 1 to 4,
the equation on the left represents the Lie derivative associated
with the metric tensor along with the vector field $\xi$. It is
important to note that the function $\Phi$ acts as an arbitrary
function of the radial $r$ as well as time $t$ coordinate, even in
the static scenario. Now, by applying Eq. (\ref{Killing}) on the
spacetime via Eq. (\ref{eq:13a}), the following form of conformal
Killing equations can be obtained:
\begin{eqnarray}\label{Killing1}
\xi^1 a'=\Psi,~~~~~ \xi^4=W,~~~~~\xi^1=\frac{\Psi~r}{2},~~~~~\xi^1
b'+2{\xi^1}_1=\Psi,
\end{eqnarray}
that lead to a simultaneous solution given by
\begin{eqnarray}\label{Killing2}
e^a=X^2 r^2,~~~~~
e^b=\big(\frac{Y}{\Psi}\big)^2,~~~~~\xi^i=W~{\delta^i_4}
+\big(\frac{r\Psi}{2}\big){\delta^i_1},
\end{eqnarray}
in which $W$, $X$ and $Y$ are considered as the arbitrary constants.
Now, by plugging Eq. (\ref{Killing2}) into Eqs.
(\ref{eq:24}-\ref{eq:26}), we have the following field equations
under the effect of conformal symmetry:
\begin{eqnarray}
    8\pi\rho +E^{2} &&= -\frac{\alpha  \Psi (r)^2}{Y ^2 r^2}+\frac{\alpha }{r^2}-\frac{2 \alpha  \Psi (r) \Psi '(r)}{Y ^2 r}+\frac{\Phi }{2},\label{eq:27}\\
    8\pi p -E^{2} &&= \frac{3 \alpha  \Psi (r)^2}{Y ^2 r^2}-\frac{\alpha }{r^2}-\frac{\Phi }{2},\label{eq:28}\\
    8\pi p +E^{2} &&= \frac{\alpha  \Psi (r)^2}{Y ^2 r^2}+\frac{2 \alpha  \Psi (r) \Psi '(r)}{Y ^2 r}-\frac{\Phi }{2},\\\label{eq:29}
\end{eqnarray}
On solving this field equation simultaneously, the expressions for
the energy density, pressure, and  electric field can be attained as
follows:
\begin{eqnarray}
 \rho &&= \frac{\alpha }{16 \pi  r^2}-\frac{3 \alpha  \Psi (r) \Psi '(r)}{8 \pi  Y ^2 r}+\frac{\Phi }{16 \pi },\label{eq:30}\\
 p &&= \frac{\alpha  \Psi (r)^2}{4 \pi  Y ^2 r^2}-\frac{\alpha }{16 \pi  r^2}+\frac{\alpha  \Psi (r) \Psi '(r)}{8 \pi  Y ^2 r}-\frac{\Phi }{16 \pi },\label{eq:31}\\
    E^{2} &&= -\frac{\alpha  \Psi (r)^2}{Y ^2 r^2}+\frac{\alpha }{2 r^2}+\frac{\alpha  \Psi (r) \Psi '(r)}{Y ^2 r},\\\label{eq:32}
\end{eqnarray}

\subsection{Interior region of the charged gravastar admitting conformal motion}

The connection between the physical factors and the metric
potentials from Eqs. (\ref{eq:30}) and (\ref{eq:31}) is provide as:
\begin{eqnarray}\label{eq:33}
8\pi(\rho+p)=\frac{\alpha \Psi (r) \left(\psi (r)-r\Psi
'(r)\right)}{4 \pi  Y ^2 r^2}
\end{eqnarray}
Now, by using the ansatz $\rho+p=0$, we get the exact expression for
$\Psi (r)$ from Eq. (\ref{eq:33}), which is calculated as:
\begin{eqnarray}\label{eq:34}
\Psi (r)=r\Psi_{0}  ,\;\;\;\;\;\Psi (r)=0,
\end{eqnarray}
where $\Psi_{0}$ is a constant of integration. In the above
equation, $\psi (r)=0$ is not a physical solution. Now, by plugging
the values of $\psi (r)$, the physical parameters are obtained as:
\begin{eqnarray}\label{eq:35}
 \rho= \frac{-\frac{6 \alpha  \Psi_{0}}{Y ^2}+\frac{\alpha }{r^2}+\Phi }{16 \pi }=-p,\quad
 E^{2} =\frac{\alpha }{2 r^2},\quad
\sigma =\frac{\psi _0 \sqrt{\frac{\alpha }{r^2}}}{4 \sqrt{2} \pi  Y
},
\end{eqnarray}

The respective expressions of lapse functions become
\begin{equation}\label{20}
e^{a}=X^{2}r^{2}, \quad e^{-b}=\frac{r^2 \psi_0^2}{Y ^2}.
\end{equation}

The active mass of gravitation $M(r)$, within the scope of the field
Eq. (\ref{eq:30}) is given by:
\begin{eqnarray}\label{eq:38}
M(r)=4 \pi  \int_0^r r^2 \left(\rho +\frac{ E^{2} }{8 \pi }\right)
\, dr = 4 \pi  \left(-\frac{\alpha  r^3 \psi _0^2}{8 \pi  Y
^2}+\frac{r^3 \Phi }{48 \pi }+\frac{\alpha  r}{8 \pi }\right).
\end{eqnarray}

\subsection{Shell of the charged gravastar admitting conformal motion}

Herein, we shall calculate the exact solutions for the physical
parameters in the framework of EoS $p=\rho$ by using Eq.
(\ref{eq:30}) and Eq. (\ref{eq:31}). The EoS $p=\rho$ is calculate
as
\begin{eqnarray}\label{eq:39}
\frac{Y ^2 \left(\alpha +r^2 \Phi \right)-6 \alpha  r\Psi (r)\Psi
'(r)}{16 \pi  Y ^2 r^2}=\frac{-Y ^2 \left(\alpha +r^2 \Phi \right)+2
\alpha  r\Psi (r)\Psi '(r)+4 \alpha \Psi (r)^2}{16 \pi  Y ^2 r^2}.
\end{eqnarray}
Now, by employing the EoS $p=\rho$, the exact solution for $\Psi
(r)$ from Eq. (\ref{eq:39}) is calculated as:
\begin{eqnarray}\label{eq:40}
\psi=\pm\frac{\sqrt{6 \alpha \Psi_{1}+Y ^2 r \left(3 \alpha +r^2
\Phi \right)}}{\sqrt{6} \sqrt{\alpha } \sqrt{r}}.
\end{eqnarray}
with $\Psi_{1}$ as an integration constant. In the above equation,
$\psi (r)=0$ is not a physical solution. Now, by plugging the values
of $\psi (r)$, we obtain the expressions of the physical parameters
as:
\begin{eqnarray}\label{eq:41}
 \rho = \frac{\alpha  \left(3\Psi_{1}+Y ^2 r\right)}{16 \pi  Y ^2 r^3}=p,\quad
 E^{2}=-\frac{3 \alpha \Psi_{1}}{2 Y ^2 r^3},\quad
\sigma =\frac{\sqrt{-\frac{\alpha \Psi_{1}}{Y ^2 r^3}}
\sqrt{\frac{6\Psi_{1}}{Y ^2 r}+\frac{r^2 \Phi }{\alpha }+3}}{16 \pi
r},
\end{eqnarray}

The respective expressions of lapse functions become
\begin{equation}\label{20}
e^{a}=X^{2}r^{2}, \quad e^{-b}=\frac{6 \alpha  \Psi_{1}+Y ^2 r
\left(3 \alpha +r^2 \Phi \right)}{6 \alpha  Y ^2 r}.
\end{equation}

The active mass of gravitation $M(r)$ within Eq. (\ref{eq:30}) is
given by
\begin{eqnarray}\label{eq:44}
M(r)=4 \pi  \int_0^r r^2 \left(\rho +\frac{ E^{2}}{8 \pi }\right) \,
dr = \frac{\alpha  r}{4}.
\end{eqnarray}

\subsection{Exterior region of the charged gravastar admitting conformal motion}

Herein, we shall calculate the exact solutions for physical
parameters in the framework of EoS $p=\omega\rho$ with $\omega=0$ by
using Eq. (\ref{eq:30}) and Eq. (\ref{eq:31}). Now, implementing the
EoS, the exact solution for $\Psi (r)$ from Eq.
(\ref{eq:39}) is calculated as:
\begin{eqnarray}\label{eq:40}
\psi=\pm\frac{\sqrt{12 \alpha  \Psi_{2}+Y ^2 r^4 \left(3 \alpha +2
r^2 \Phi \right)}}{2 \sqrt{3} \sqrt{\alpha } r^2}.
\end{eqnarray}
where $\Psi_{2}$ is a constant of integration. In the above
equation, $\psi (r)=0$ is not a physical solution. Now, by plugging
the values of $\psi (r)$, we get the following expressions for the
physical parameters:
\begin{eqnarray}\label{eq:42}
E^{2}=\frac{\alpha  \left(r^4-\frac{12 \Psi_{2}}{Y ^2}\right)}{4
r^6},\quad \sigma =\frac{\alpha \left(12\Psi_{2}+Y ^2 r^4\right)
\sqrt{\frac{12 \Psi_{2}}{Y ^2 r^4}+\frac{2 r^2 \Phi }{\alpha
}+3}}{16 \sqrt{3} \pi  Y ^2 r^4 \sqrt{\alpha \left(r^4-\frac{12
\Psi_{2}}{Y ^2}\right)}},
\end{eqnarray}

The respective expressions of lapse functions become
\begin{equation}\label{20}
e^{a}=X^{2}r^{2}, \quad e^{-b}=\frac{12 \alpha  \Psi_{2}+Y ^2 r^4
\left(3 \alpha +2 r^2 \Phi \right)}{12 \alpha  Y ^2 r^4}.
\end{equation}

Now, to manifest that the exterior region is a flat
geometry, we derive the Kretschmann scalar $(K_S)$ which is defined
as
\begin{eqnarray}\label{eq:44}
K_S= R^{ijkl}R_{ijkl},
\end{eqnarray}
in which $R$ shows the Riemann tensor. The $K_S$ corresponding to
the exterior geometry leads to,

\begin{eqnarray}\label{eq:44}
K_S=\frac{\left(12 \alpha  \Psi_{2}+Y ^2 r^4 \left(3 \alpha +2 r^2
\Phi \right)\right){}^2 \left(1872 \alpha ^2 \Psi_{2}^2+24 \alpha
\Psi_{2} Y ^2 r^4 \left(3 \alpha -10 r^2 \Phi \right)+Y ^4 r^8
\left(45 \alpha ^2+52 r^4 \Phi ^2+84 \alpha r^2 \Phi
\right)\right)}{432 \alpha ^4 Y ^8 r^{20}},
\end{eqnarray}
It is very interesting to mention that $K_S\rightarrow 0$ as
$r\rightarrow \infty$ by assuming $\Phi=0$. This shows that the
exterior geometry is asymptotically flat spacetime.

\subsection{Boundary constraints}

Boundary conditions play an important role in computing the values of
different constants that appear in the system. We now estimate the
constant A and examine its dependency on charge $q$ by equating the
interior and thin shell domains at the boundary $r = r_1$. Also, we
evaluate the range of the outer radius as well as the thickness ($r_2 -
r_1$) of the thin shell by matching the thin shell region with the
exterior domain at $r = r_2$. Here, we adopt the interior radius as
$R_1 = 10km$ \cite{44}.

\begin{itemize}

\item  Equating the regions of interior and thin shell at $r = r_1 = 10 km$, we obtain:
\begin{eqnarray}\label{eq:44}
X^{2}r^{2}=X^{2}r^{2},
\end{eqnarray}
for $g_{tt}$ component and also for $g_{rr}$ component, we have
\begin{eqnarray}\label{eq:44}
\frac{r_1^2 \psi _0^2}{Y ^2}=\frac{6 \alpha  \Psi_{2}+Y ^2 r \left(3
\alpha +r_1^2 \Phi \right)}{6 \alpha  Y ^2 r_1},
\end{eqnarray}

\item Also, by matching the thin shell with the exterior domain at $r = r_2$, we obtain:
\begin{eqnarray}\label{eq:44}
X^{2} r_2^{2}=X^{2} r_2^{2},
\end{eqnarray}
for $g_{tt}$ component and also for $g_{rr}$  component, we have
\begin{eqnarray}\label{eq:44}
\frac{6 \alpha  \Psi_{1}+Y ^2 r_2 \left(3 \alpha + r_2^2 \Phi
\right)}{6 \alpha  Y ^2  r_2}=\frac{12 \alpha \Psi_{2}+Y ^2 r_2^4
\left(3 \alpha +2  r_2^2 \Phi \right)}{12 \alpha  Y ^2 r_2^4},
\end{eqnarray}

\end{itemize}

\subsection{Junction conditions}

In this section, we match inner and outer developed solutions through
well-known Darmoise-Israel formalism. We have studied that the
metric coefficients are continuous across the junction while
calculating the boundary conditions. Whether their derivatives are
also continuous or not is subject to investigation. We match these
spacetimes at the hypersurface $(\Omega)$, i.e., at $r=z $. With the
help of the Lanczos equation, the surface stress-energy tensor is
computed as:
\begin{equation}\label{25}
S_{\beta}^{\alpha}=\frac{1}{8\pi}(\delta_{\beta}^{\alpha}
\zeta_{\gamma}^{\gamma}-\zeta_{\beta}^{\alpha}),
\end{equation}
where $\zeta_{\alpha\beta}=K^{+}_{\alpha\beta}-K^{-}_{\alpha\beta}$
which exhibits the discontinuous nature of the extrinsic curvature.
The negative and positive signatures manifest the internal and
external eras, respectively. The constituents of extrinsic curvature
at $\Omega$ are provided by:
\begin{equation}\label{26}
{K_{\alpha\beta}^{\pm}}= -n_{\mu}^\pm \left[\frac{\partial^2
x^{\mu}}{\partial \eta^{\alpha}
\eta^{\beta}}+\Gamma^{\mu}_{Y\nu}\left(\frac{\partial
x^{Y}}{\partial\eta^{\alpha}}\right)\left(\frac{\partial
x^{\nu}}{\partial\eta^{\beta}}\right)\right],
\end{equation}
where $\eta^{\beta}$ depicts the coordinates of the internal shell
while $n_{\mu}^{\pm}$ describes the unit normal at $\Omega$ given by
\begin{equation}\label{27}
n_{\mu}^{\pm}=\pm\left|g^{\beta\nu}\frac{\partial L}{\partial
x^{\beta}}\frac{\partial L}{\partial
x^{\nu}}\right|^{-\frac{1}{2}}\frac{\partial L}{\partial x^{\mu}},
\quad n_{\mu} n^{\mu}=1.
\end{equation}

Corresponding to the perfect fluid configuration, we get
$S_{\beta}^{\alpha}= \text{diag}(\varrho, -\mathbb{P}, -\mathbb{P})$
in which $\mathbb{P}$ and $\varrho$ are surface pressure and energy
density, respectively that are illustrated via Lanczos equations as
\begin{eqnarray}\label{28}
\varrho=-\frac{1}{4\pi
\mathcal{R}}\left[\sqrt{L}\right]_{-}^{+},\quad \mathbb{P}=
-\frac{\varrho}{2}+\frac{1}{16\pi}\left[\frac{{L^{'}}}
{\sqrt{L}}\right]_{-}^{+}.
\end{eqnarray}

In the present research, we consider two choices for the exterior
manifolds.

\begin{itemize}
\item First, we consider the metric components of internal and external manifolds that are calculated through the field equations with boundary conditions.
\begin{eqnarray}\label{28}
L_-=\frac{z \left(-3 \alpha  Y ^2+6 \alpha  z^2 \psi_0^2-Y ^2 z^2
\Phi \right)}{6 \alpha },\quad L_+=\frac{Y ^2 z^4}{4-4 z^3}.
\end{eqnarray}

On inserting the metric potentials of internal and external
manifolds in the above equations, the matter variables are derived
as
\begin{eqnarray}\label{30}
\varrho_b&=&-\frac{1}{4 \pi z}\left[\sqrt{\frac{Y ^2 z^4}{4-4
z^3}}-\sqrt{\frac{z \left(-3 \alpha  Y ^2+6 \alpha z^2 \psi_0^2-Y ^2
z^2 \Phi \right)}{6 \alpha }}\right],
\\\label{31} \mathbb{P}_b&=&\frac{1}{8 \pi z}\left(\frac{3
\left(z^3-2\right) \left(-\frac{Y ^2 z^4}{z^3-1}\right)^{3/2}}{2 Y
^2 z^4}+\frac{-5 z^3 \left(Y ^2 \Phi -6 \alpha \psi_0^2\right)-9
\alpha Y ^2 z}{\alpha  \sqrt{36 z^3 \psi_0^2-\frac{6 Y ^2z \left(3
\alpha +z^2 \Phi \right)}{\alpha }}}\right).
\end{eqnarray}

\item Secondly, we consider RN BH as an exterior manifold and inner spacetime is considered a calculated solution
through field equations given as
\begin{eqnarray}\label{28}
L_-=\frac{z^2 \psi_0^2}{Y ^2},\quad L_+=-\frac{2
m}{z}+\frac{Q^2}{z^2}+1.
\end{eqnarray}

On inserting the metric potentials of internal and external
manifolds in the above equations, the matter variables are derived
as
\begin{eqnarray}\label{30}
\varrho_b&=&\frac{1}{4 \pi  z}\left[\sqrt{\frac{z^2 \psi _0^2}{Y
^2}}-\sqrt{\frac{-2 m z+Q^2+z^2}{z^2}}\right],
\\\label{31} \mathbb{P}_b&=&\frac{1}{8 \pi
z}\left[\frac{2 (z-m)}{z \sqrt{\frac{-2 m z+Q^2+z^2}{z^2}}}-4
\sqrt{\frac{z^2 \Psi_{0}}{Y ^2}}\right].
\end{eqnarray}

\end{itemize}

From surface energy density, the mass of the shell is calculated as
\begin{equation}\nonumber
\mathcal{M}_{shell}=4\pi z^{2}\varrho.
\end{equation}

\section{Some Physical Attributes of Charged Gravastars}

This section is devoted to analyzing the some of physical attributes
of charged gravastars such as the EoS parameter, the proper length,
energy and entropy admitting the CKVs. First, we are going to
explore the EoS parameter.

\subsection{The EoS Parameter}

It is well known that the connections among the matter variables
such as the pressure and energy density can be efficiently
demonstrated by the EoS parameter. At $r=z$, the EoS parameter
associated with the matter constituents is displayed as
\begin{equation}\label{33}
\zeta(z)=\mathbb{P}/\varrho .
\end{equation}

\begin{itemize}
\item For the first case: the substitution of the associated values of matter contents provides
\begin{equation}\label{33}
\zeta(z)=\frac{3 \left(\frac{3 \left(z^3-2\right) \left(-\frac{Y ^2
z^4}{z^3-1}\right)^{3/2}}{2 Y ^2 z^4}+\frac{-5 z^3 \left(Y ^2 \Phi
-6 \alpha  \psi _0^2\right)-9 \alpha  Y ^2 z}{\alpha \sqrt{36 z^3
\psi _0^2-\frac{6 Y ^2 z \left(3 \alpha +z^2 \Phi \right)}{\alpha
}}}\right)}{-\sqrt{36 z^3 \psi_0^2-\frac{6 Y ^2 z \left(3 \alpha
+z^2 \Phi \right)}{\alpha }}+3 \sqrt{-\frac{Y ^2 z^4}{z^3-1}}}.
\end{equation}

The sensitivity of this equation is enhanced due to the existence of
several terms with square roots and fractions. The fundamental
constraints for real EoS parameters are given as

\begin{eqnarray}\nonumber
&&\psi_0<0,\Phi \leq 0,\alpha >-\frac{1}{3} \left(z^2 \Phi
\right)\land -\sqrt{6} \sqrt{\frac{\alpha  z^2 \psi_0^2}{3 \alpha
+z^2 \Phi }}<Y <\sqrt{6} \sqrt{\frac{\alpha  z^2 \psi_0^2}{3 \alpha
+z^2 \Phi }},z>0,\\\nonumber &&\psi_0<0,\Phi
>0,\alpha
>0\land -\sqrt{6} \sqrt{\frac{\alpha  z^2 \psi_0^2}{3 \alpha +z^2
\Phi }}<Y <\sqrt{6} \sqrt{\frac{\alpha z^2 \psi_0^2}{3 \alpha +z^2
\Phi }},z>0, \\\nonumber &&\psi_0>0,\Phi \leq 0,\alpha
>-\frac{1}{3} \left(z^2 \Phi \right)\land -\sqrt{6}
\sqrt{\frac{\alpha  z^2 \psi_0^2}{3 \alpha +z^2 \Phi }}<Y <\sqrt{6}
\sqrt{\frac{\alpha z^2 \psi_0^2}{3 \alpha +z^2 \Phi
}},z>0,\\\nonumber &&\psi_0>0,\Phi
>0,\alpha >0\land -\sqrt{6} \sqrt{\frac{\alpha  z^2 \psi_0^2}{3
\alpha +z^2 \Phi }}<Y <\sqrt{6} \sqrt{\frac{\alpha z^2 \psi_0^2}{3
\alpha +z^2 \Phi }},z>0,
\end{eqnarray}

\item For the second case: The substitution of the associated values of matter contents yields
\begin{equation}\label{33}
\zeta(z)=\frac{\frac{2 (z-m)}{z \sqrt{\frac{-2 m
z+Q^2+z^2}{z^2}}}-4 \sqrt{\frac{z^2\psi_0^2}{Y ^2}}}{2
\left(\sqrt{\frac{z^2 \psi_0^2}{Y ^2}}-\sqrt{\frac{-2 m
z+Q^2+z^2}{z^2}}\right)}.
\end{equation}

The fundamental constraints for real EoS in this case are given by

\begin{eqnarray}\nonumber
&&m<\frac{z}{3}\land \left(-\sqrt{\frac{-2 Y ^2 m z+Y ^2 Q^2+Y ^2
z^2}{z^4}}<\text{$\psi _0$}<-\frac{1}{2} \sqrt{\frac{Y ^2 m^2-2 Y ^2
m z+Y ^2 z^2}{z^2 \left(-2 m
z+Q^2+z^2\right)}}\right.\\\nonumber&&\left.\lor \frac{1}{2}
\sqrt{\frac{Y ^2 m^2-2 Y ^2 m z+Y ^2 z^2}{z^2 \left(-2 m
z+Q^2+z^2\right)}}<\psi _0<\sqrt{\frac{-2 Y ^2 m z+Y ^2 Q^2+Y ^2
z^2}{z^4}}\right),\\\nonumber &&
\frac{r}{3}<m<\frac{z}{2},Q>\frac{\sqrt{3 m z-z^2}}{\sqrt{2}}\land
\left(-\sqrt{\frac{-2 Y ^2 m z+Y ^2 Q^2+Y ^2 z^2}{z^4}}<\psi_0
<-\frac{1}{2} \sqrt{\frac{Y ^2 m^2-2 Y ^2 m z+Y ^2 z^2}{z^2 \left(-2
m z+Q^2+z^2\right)}}\right.\\\nonumber&&\left.\lor \frac{1}{2}
\sqrt{\frac{Y ^2 m^2-2 Y ^2 m z+Y ^2z^2}{z^2 \left(-2 m
z+Q^2+z^2\right)}}<\psi _0<\sqrt{\frac{-2 Y ^2 m r+Y ^2 Q^2+Y ^2
z^2}{z^4}}\right).
\end{eqnarray}

These mathematical expressions of $z$ indicate the geometry of gravastar-like compact structures which are similar to a dust shell without pressure component. In the exterior region, two methodologies are employed: firstly, calculating the vacuum exact solution, and secondly, utilizing the Reissner-Nordström metric. These spacetimes are matched through junction conditions, and stability constraints are found for both cases. It is clearly shown through the constraints that the impact of different exterior choices greatly effect the stable configurations. It is found that the conformal constants and shell radius play a remarkable role in maintaining the stable structure of the developed gravastars.

\end{itemize}

\subsection{Proper length}

\begin{figure}\centering
\epsfig{file=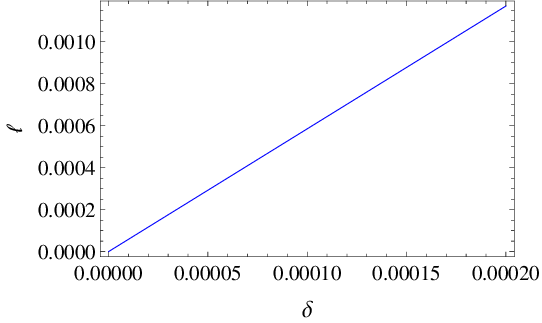,width=.33\linewidth}\epsfig{file=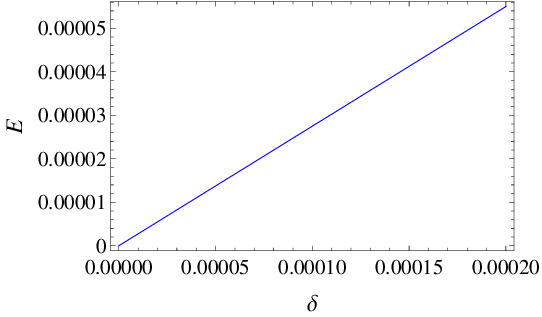,width=.33\linewidth}\epsfig{file=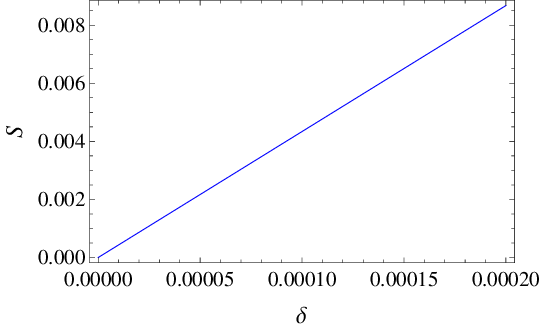,width=.33\linewidth}
\caption{\label{p} Proper length (left plot), energy contents
(middle plot) and entropy of shell (right plot) for
$\Psi_{1}=1,\alpha =0.5,Y =0.5,\Phi =1,z=10;$.}
\end{figure}

We are interested to discuss the proper length of the shell and the
thickness of the shell is represented by $\delta$. The shell
thickness is a very small positive real number such as
$0<\delta\ll1$. The lower and upper boundaries of the shell are $z$
and $z+\delta$. Mathematically, the proper length of the shell can
be evaluated as \cite{qm}
\begin{equation}\label{19}
l=\int_{z}^{z+\delta}\sqrt{e^{b(r)}}=\int_{z}^{z+\delta}
\sqrt{\frac{6 \alpha  \Psi_{1}+Y ^2 r \left(3 \alpha +r^2 \Phi
\right)}{6 \alpha  Y ^2 r}}dr.
\end{equation}
To solve the above-complicated integration, we assume that
the $\sqrt{\frac{6 \alpha  \Psi_{1}+Y ^2 r \left(3 \alpha +r^2 \Phi
\right)}{6 \alpha  Y ^2 r}}=\frac{dB(r)}{dr}$ as
\begin{equation}\label{20}
l=\int_{z}^{z+\delta}\frac{dB(r)}{dr}dr=B(z+\delta)-B(z)\approx
\delta \frac{dB(r)}{dr}|_{r=z}=\delta\sqrt{\frac{6 \alpha \Psi_{1}+Y
^2 z \left(3 \alpha +z^2 \Phi \right)}{6 \alpha Y ^2 z}}.
\end{equation}
The square and higher powers of $deltall1$, a tiny positive real
constant, can be ignored. As a consequence, a correlation between
the shell's thickness and appropriate length can be studied. The
physical parameters and the shell radius both affect this
connection. The correct length varies depending on the thickness of
the shell as shown by the left plot in Fig. (\ref{p}). The suitable
length grows together with the thickness for every choice of
physical parameters.

\subsection{Energy content}

In the inner area of a gravastar, where matter obeys the equation of
state $p=-\rho$, there is a negative energy zone and a
non-attractive force. Analyzing the energy distribution within the
shell region can be done using a similar method, such as figuring
out the right length as \cite{qm}
\begin{equation}\label{25}
\textbf{E}=\int_{z}^{z+\delta}4\pi r^2 \rho(r)dr=\frac{1}{4} \alpha
\left(\frac{3 \Psi_{1} (\log (\delta +z)-\log (z))}{Y ^2}+\delta
\right).
\end{equation}
The shell energy's final expression depends on the shells
thickness, shell's radius, $\Phi$ and $\alpha$. The middle plot of Fig.
(\ref{p}) shows the behavior of shell energy along the thickness of
the shell for suitable values of physical parameters. It is noted
that shell energy increases as the thickness of the shell increases.

\subsection{Entropy}

A geometric structure's entropy value can be used to calculate the
degree of disorder or disturbance in the structure. The entropy of
thin-shell gravastars are investigated to quantify the
unpredictability of gravastar geometry. The idea put out by Mazur
and Mottola is used to create an equation for a thin-shell
gravastar's entropy as \cite{qm}
\begin{equation}\label{22}
S=\int_{z}^{z+\delta}4\pi r^2 j(r) \sqrt{e^{\varepsilon(r)}}dr.
\end{equation}
For local temperature, the entropy density is calculated as
\begin{equation}\label{23}
j(r)=\frac{\eta K_B}{\hbar}\sqrt{\frac{p(r)}{2\pi}},
\end{equation}
where $\eta$ is represented as a dimensionless parameter. Here, we take
Planck units $(K_B = 1 = \hbar)$ so that the shell's entropy becomes
\cite{qm}
\begin{equation}\label{24}
S= \delta \frac{z^2 \sqrt{\frac{\alpha  \left(\frac{3 \Psi_{1}}{Y
^2}+z\right)}{z^3}} \sqrt{\frac{6 \Psi_{1}}{Y ^2 z}+\frac{z^2 \Phi
}{\alpha }+3}}{2 \sqrt{3}}.
\end{equation}
Also, the shell's entropy must be proportional to $\delta$. It is
noted the shell entropy also enhances as the shell's thickness
increases. The shell's entropy approaches zero as the thickness of the shell
approaches to zero as shown in the right plot of Fig. (\ref{p}).

\section{Conclusion}\label{sec:6}

A compact, spherically symmetric astrophysical object called a "gravastar" provides a potential solution to the black hole geometry's singularity problem. If gravastar exists and can be found in our universe, that is the question at discussion. There is debate about whether the gravitational waves recorded by LIGO are the consequence of merging gravastars or black holes, even though there is no scientific evidence in favor of gravastars. The theoretical existence of gravastar and its physical viability are shown in this article. In contrast to previous similar efforts, this study's novel implications come from investigating charge effects on gravastars in the setting of $f(Q)$ gravity. Prior studies might have concentrated on gravastar characteristics without taking charge or particular gravity models such as $f(Q)$ into account. This work illuminates how the interaction of charge and gravity shapes the stability and physical properties of gravastars by introducing the notion of charge and applying a specific gravity model. Furthermore, the comprehensive understanding of charged gravastar solutions that was previously lacking in the literature is provided by the detailed analysis of the gravastar's internal, intermediate, and external regions as well as the consideration of various spacetime solutions and stability constraints. Therefore, this study's distinctive contributions include a more comprehensive investigation of charged gravastars and their properties compared to previous studies that might have concentrated on various areas or lacked the depth of analysis that this study provides. In the current study, we are interested in creating charged gravastar solutions utilizing conformal motion within the context of symmetric teleparallel gravity by using the particular model $f(Q)=\alpha  Q+\Phi$. Gravastar geometry may be divided into three sections known as internal, intermediate, and exterior regions, each having a distinct form of EoS as $p=\omega\rho$ and $\omega=-1,1,0$, respectively. Then, we took into account these EoS and conformal Killing vectors to determine the precise solutions for each of these locations. It should be emphasized that each computed solution is regular and also confirmed by the Kretschmann scalar. Then, using junction conditions, we connected these sections to build the geometry of gravastars. We took two scenarios for the external geometry into consideration here. To create the gravastars' geometry, we first employed the computed exterior solution. To examine the configurations of gravastars, we also treated the RN black hole as an outside manifold.

Then, we explored some physical characteristics of charged gravastars. By using the EoS parameter which is also referred to as the speed of sound parameter, we discussed the stability for both cases as exterior obtained solution and assumed RN black hole. Then, we calculated the stability constraints that depend on the coupling constant as well as conformal parameters. The proper length, energy content and entropy of the developed structure are evaluated. It is interesting to mention that these physical attributes are directly proportional to the thickness of the shell. 

\section*{Acknowledgement}

F. Javed acknowledges the financial support provided through Grant
No. YS304023917, which has contributed to his Postdoctoral
Fellowship at Zhejiang Normal University. This research is partly supported by Research Grant F-FA-2021-510 of the Uzbekistan Ministry for Innovative Development. This research was supported by the researchers Supporting Project Number (RSP2024R401), King Saud University, Riyadh, Saudi Arabia.

\end{document}